\documentclass[12pt]{article}
\usepackage{moriond,epsfig}
\usepackage[latin1]{inputenc}
\usepackage[OT1]{fontenc}

\def\be{\begin{equation}}
\def\ee{\end{equation}}
\def\bea{\begin{eqnarray}}
\def\eea{\end{eqnarray}}

\def\cF{{\cal F}}

\def\cG{{\cal G}}
\def\cR{{\cal R}}

\newcommand{\ec}{\ensuremath{\varepsilon_{c}}}
\newcommand{\ve}{\varepsilon}
\newcommand{\bei}{\begin{itemize}}
\newcommand{\eei}{\end{itemize}}
\newcommand{\non}{\nonumber}
\begin{document}
\vspace*{4cm}
\title{NON-ABELIAN CONDENSATES AS ALTERNATIVE FOR DARK ENERGY \footnote{
Extended version of the talk given at 43-rd Rencontres de Moriond,
La Thuile, March 15-22, 2008}}

\author{DMITRI V. GAL'TSOV }

\address{Department of Physics, Moscow State University, Russia} \maketitle

\baselineskip=19pt \abstracts{ We review basic features of
cosmological models with large-scale classical non-Abelian
Yang-Mills (YM) condensates. There exists a unique $SU(2)$ YM
configuration (generalizable to larger gauge groups) compatible with
homogeneity and isotropy of the three-space which is parameterized
by a single scalar field. In the past various aspects of
Einstein-Yang-Mills (EYM) cosmology were discussed in the context of
the Early Universe. Due to conformal invariance, solvable EYM FRW
models exist both on the classical and quantum levels. To develop
the YM model for dark energy one has to find mechanisms of the
conformal symmetry breaking. We discuss the Born-Infeld
generalization and some phenomenological models motivated by quantum
corrections  exploring possibility of transient DE and phantom
regimes.}

\section{Introduction}
Vector fields as the dark energy candidates were recently proposed
by several
authors~\cite{ArmendarizPicon:2004pm,Kiselev:2004py,Zhao:2005bu,Wei:2006tn,Fuzfa:2006pn,Zhao:2007vn,Jimenez:2008au,Bamba:2008xa}.
An attractive feature of this suggestion is the fact that vector
fields are the basic constituents of the standard model and its
generalizations, while the scalar fields invoked in the most popular
dark energy models~\cite{Copeland:2006wr} remain rather speculative.
Formation of non-Abelian condensates in cosmology within the context
of the standard model and GUT-s may have various
reasons~\cite{Linde:1979pr}. An inflationary model driven by a
vector field was proposed by Ford~\cite{Ford:1989me} long ago.
Basically the same ideas lie behind   more recent proposals of
vector   dark energy. A major problem with  vector fields is that
they generically single out spatial directions leading to
anisotropy, unless one is just looking for the source of possible
anisotropy~\cite{Zimdahl:2000zm}. To ensure isotopy one has to
introduce suitable multiplets of vector fields. A natural framework
for this is non-Abelian gauge theory.
\par
A remarkable fact is that the $SU(2)$ YM field allows for isotropic
and homogeneous configurations parameterized by a single scalar
function. This ansatz is compatible with the FRW models of any type,
closed, open or spatially flat~\cite{Galtsov:1991un} (the
compatibility of the closed FRW model with the $SU(2)$ YM,
 quite obvious due to isomorphism $SU(2)\sim S^3$, was noticed
earlier by several
authors~\cite{Cervero:1978db,Henneaux:1982vs,Hosotani:1984wj}). This
property remains valid  also for larger gauge groups containing an
embedded
SU(2)~\cite{Bertolami:1990je,Moniz:1990hf,Moniz:1991kx,Kuenzle:1991wa,Darian:1996mb}.
Remarkably, the  EYM cosmology is also solvable at the quantum
level,  the quantum FRW cosmology having been discussed in a number
of
papers~\cite{Bertolami:1990iw,Donets:1992ck,Donets:1992vx,Cavaglia:1993en,Bertolami:1994jn,Cavaglia:1994zv}.
An interesting feature the EYM quantum cosmology is possibility of
tunneling transitions between  de Sitter and hot FRW cosmologies. An
embedding of the EYM cosmology into string theory was also
discussed~\cite{Bertolami:1991ff,Bento:1994dw,Moniz:1996ja}.
\par
Dynamics of classical YM fields in the homogeneous and isotropic FRW
cosmology is perfectly regular. In the homogeneous but anisotropic
models the YM field in contrary often exhibit chaotic
features~\cite{Darian:1996jf,Barrow:1997sb,Jin:2004vh}. This
behavior is similar to  chaotic behavior of homogeneous (depending
only on time) YM fields in flat space~\cite{Matinyan:1981dj}. A very
peculiar dynamics was observed in the Kantowski-Sachs non-Abelian
cosmology, which was investigated as an interior solution for
non-Abelian black
holes~\cite{Donets:1996ja,Donets:1998xd,Gal'tsov:1997uq,Galtsov:1997iy,Galtsov:1997ub}.
It is non-chaotic, but the approach to singularity is characterized
by infinitely growing oscillations. Other attempts to apply YM
fields in cosmology amount to the idea of growing defects, it was
applied to Einstein-Yang-Mills
sphalerons~\cite{Galtsov:1991du,Donets:1992zb}
in~\cite{Ding:1995jf}. For   more general applications of
self-gravitating YM fields see
Refs.~\cite{Volkov:1998cc,Gal'tsov:2001tx}.
\par
Dynamics of the YM fields governed by the standard action quadratic
in the field strength is conformally invariant (in four dimensions).
In the cosmological context this means that the equation of state
arising in the framework of the FRW cosmology will be that of the
photon gas $p=\epsilon/3$, therefore one actually deals with a cold
substitute for the hot  gas~\cite{Galtsov:1991un}. It was even
argued that the classical YM condensate could be part of  the
CMB~\cite{Tipler:2007vx}. Conformal invariance implies that if the
YM condensate was formed before inflation, it would be strongly
diluted during the de Sitter era. Meanwhile, conformal invariance
might be broken for various reasons, such as string theory
modifications, quantum corrections or interactions with other
fields. Within the context of string theory one deals with the
Born-Infeld type action for YM fields rather than the Maxwell
action, which is no more conformally invariant (for discussion and
references see~\cite{Gal'tsov:1999vn}). The FRW cosmology driven by
a non-Abelian Born-Infeld action was studied in detail in
Ref.~\cite{Dyadichev:2001su}. The equation of state in this model
interpolates between the string gas equation $p=-\epsilon/3$ and
that of the hot photon gas. The negative pressure may generate the
regime of zero deceleration, but it is still insufficient to ensure
cosmic acceleration. Further generalizations of this model were
considered
in~\cite{Moniz:2002rd,Moniz:2002rd,Gal'tsov:2003xm,Fuzfa:2005qn,Fuzfa:2006pn}.
More general phenomenological models with vector fields governed by
dynamics of the Born-Infeld type were also discussed
in~\cite{Novello:2006ng}. Note that in anisotropic cosmology the
Born-Infeld Lagrangian typically leads to damping off the chaotic
oscillations~\cite{Gal'tsov:2003gx,Dyadichev:2004ix,Dyadichev:2006xc}.
\par
Here we briefly discuss various aspects of cosmological models with
classical large scale YM fields which may be helpful to better
understand  the possibility of the corresponding dark energy (DE)
models. The concrete suggestions in this directions   still look
preliminary, while the idea certainly deserves further
investigation.

\section{SU(2) homogeneous and isotropic cosmology}
\subsection{YM configuration}
Space-time symmetry of the Yang-Mills field $A^a_\mu$ means that its
variation under action of the isometry transformations can be
compensated by a suitable gauge transformation. In this case the
field strength $F^a_{\mu\nu}$ transforms multiplicatively so that
the action and the stress-tensor remain invariant, implying
compatibility with the FRW space-time metric \be
ds^2=a^2(\eta)\left(d\eta^2-dl_k^2\right).\ee   Demanding such
invariance under spatial translations and $SO(3)$ rotations one
finds that the unique (up to scaling) $SU(2)$ YM configuration is
parameterized by a single scalar function of time (we use conformal
time) $w(\eta)$ and contains both electric and magnetic components:
\be E^a_i=\dot w \;\delta^a_i,\quad B^a_i= (k-w^2)\; \delta^a_i, \ee
where $k=\pm 1,0$ for closed, open and spatially flat cases (for
$k=1,0$ it was shown long ago by Cervero and
Jacobs~\cite{Cervero:1978db}, Henneaux~\cite{Henneaux:1982vs} and
Hosotani~\cite{Hosotani:1984wj}, while derivation for all the three
cases starting with the Witten ansatz was given by Gal'tsov and
Volkov~\cite{Galtsov:1991un} (for a derivation see
also~\cite{Dyadichev:2001su}). The standard YM action quadratic in
$F^a_{\mu\nu}$ reduces to the one-dimensional Lagrangian  \be
L=\frac{3}{8\pi a^4}\left({\dot w}^2-(k-w^2)^2 \right).\ee This
looks like the scalar field Lagrangian with the potential. The
kinetic term corresponds to the (color) electric field contribution,
while the potential term is due to the magnetic field. During the
evolution the electric and magnetic fields exchange the energy, and
no purely electric or purely magnetic YM fields are possible except
the static (unstable) configuration $w=0$, which is non-vacuum in
the closed case $k=1$ (the cosmological sphaleron). The dependence
$a^{-4}$ on the scale factor reflects the conformal nature of the YM
field.
\subsection{Cold matter for hot Universe}
  The standard YM action is conformally invariant, so the
above configuration   gives rise to the equation of state of the
photon gas
 \be
 p=\frac{\epsilon}3\quad {\rm with}\quad \epsilon=\frac3{8\pi a^4}
\left[{\dot w}^2+(k-w^2)^2\right].\ee  Thus one obtains the hot
Universe driven by cold matter~\cite{Galtsov:1991un}.

The YM equations reduce to an equation of motion  of a particle in
the potential well \be V= (k-w^2)^2,\ee  which in the closed case
($k=1$) is the double well potential. Two minima correspond to
neighboring topologically distinct vacua.  Depending on the
$w$-particle total energy, the field classically oscillates either
around a single vacuum, or around both of them. In the flat and open
cases the potential has one minimum at $w=0$ with $V=0$ for the flat
and $V=1$ for the open cases.  The Chern-Simons 3-form
\be\omega_3=\frac{e^2}{8\pi^2} {\rm Tr} \left( A\wedge dA
-\frac{2ie}3 A\wedge A\wedge A \right),\ee satisfying the equation
\be d\omega_3= \frac{e^2}{8\pi^2} {\rm Tr} F\wedge F,\ee is
non-trivial in the case $k=1$. The Chern-Simons  number
 is equal to \be N_{CS}=\int_{S^3} \omega_3=\frac14 (w+1)^2(2-w).\ee
The vacuum $w=-1$ is topologically trivial: $N_{CS}=0$, the vacuum
$w=1$ is non-trivial: $N_{CS}=1$. Due to the axial anomaly, the
evolution of the Chern-Simons density will be accompanied by the
evolution of the fermion number density for fermions interacting
with the YM field.

\subsection{Cosmological sphaleron}
In the closed case there exists a particularly simple configuration
$w=0$ ($w$-particle sitting at the top of the barrier between two
vacua) which in analogy with  sphalerons in the Weinberg-Salam (WS)
theory was called the ``cosmological
sphaleron''~\cite{Gibbons:1993pq}. I is worth noting that the
localized particle-like EYM solutions similar to the WS sphalerons
exist as well~\cite{Galtsov:1991du} which are asymptotically flat
regular particle-like solutions of the EYM equations discovered by
Bartnik and McKinnon~\cite{Bartnik:1988am}. The repulsive YM
stresses in these objects are compensated by gravity instead of the
Higgs field in the WS sphalerons. Creation and decay of   sphalerons
generates a transition of the YM field between topological sectors,
and it is accompanied by the fermion number non-conservation in
presence of fermions~\cite{Volkov:1993gp,Volkov:1996hm}. The
cosmological sphaleron has the topological charge $N_{CS}=1/2$ like
the sphaleron  in the Weinberg-Salam theory.

The equation of motion of the $w$-particle \be  \ddot w=2w(1-w^2)
\ee is solved indeed by $w=0$, this correspond to the  total energy
\be {\dot w}^2+(1-w^2)^2=1.\ee The YM field in this case is purely
magnetic. A more general solution with the same energy describes
rolling down of the $w$-particle (the sphaleron
decay)~\cite{Gibbons:1993pq}: \be
w=\frac{\sqrt{2}}{\cosh(\sqrt{2}\eta)}.\ee  Rolling down to the
vacuum $w=1$ takes an infinite time, while the corresponding full
cosmological evolution is given by \be a=\sqrt{\frac{4\pi G}{g^2}}
\sin\eta \ee and takes a finite time. Thus the cosmological
sphaleron is quasi-stable. This conclusion is not modified if a
positive cosmological constant is added~\cite{Ding:1994nw}.
\subsection{Instantons and wormholes}
An homogeneous and isotropic EYM system has interesting features
also in the $k=1$ space of Euclidean signature which is invoked in
the path-integral formulation of quantum gravity. Actually, when
$|w|<1$, transitions between two topological sectors can be effected
via underbarrier tunneling described by instanton and wormhole
Euclidean solutions. In the Euclidean regime   the first integral of
the equations of motion reads: \be\dot w^2-(w^2-1)^2=-C, \ee where
$C$ is an integration constant. Instanton corresponds to $C=0$, it
describes tunneling between the vacua $w=\pm 1$. It is a self-dual
Euclidean YM configuration for which the stress tensor is zero,
therefore a conformally flat gravitational field can be added just
as a background.

Tunneling solutions at higher excitation levels $0<C\leq 1$ are not
self-dual. In flat space-time they are known as a meron ($C=1$, the
Euclidean counterpart of the cosmological sphaleron, $N_{CS}=1/2$)
and nested merons $0<C< 1, \;1/2 <N_{CS}<1$.  The energy-momentum
tensor of the meron is non-zero, and in flat space this solution is
singular. When gravity is added, the singularity at the location of
a meron expands to a wormhole throat, and consequently, the
Euclidean topology of the space-tme transforms to that of a
wormhole. Topological charge of the meron wormholes is zero, the
charge of the meron being swallowed by the
wormhole~\cite{Hosoya:1989zn,Das:1989ne,Rey:1989th,Gupta:1989bs,Verbin:1989sg,Donets:1992ck,Donets:1992vx}.
 The total action of these wormholes diverges because of slow
fall-off of the meron field at infinity, so the amplitude of
creation of  baby universes associated with the Euclidean wormholes
is zero. However, when a positive cosmological constant is added
(inflation) the action becomes finite due to compactness of the
space. Such solutions can be interpreted as describing tunneling
between the de Sitter space and the hot FRW universe.

  Adding the positive cosmological constant, we will get similar
first integrals both for the $w$-particle and the cosmological
radius: \be\dot w^2-(w^2-1)^2=-C,\quad \dot a^2+(\Lambda
a^4/3-a^2)=-C/(e^2 m_{Pl}^2).\ee Solutions describe independent
tunneling of   $w$  and   $a$ with different periods $T_w,\; T_a$
depending on the excitation level $C$. To be wormholes, they must
obey a quantization condition $n_w T_w=n_a T_a$~\cite{Verbin:1989sg}
with two integers. However, for a specific value of the cosmological
constant \be\Lambda=\frac34m_{Pl}^2, \ee it was
found~\cite{Donets:1992ck} that $T_a=T_w$ for all $C\in [0,1]$. In
particular, for $C=1$ (the meron limit) the radius $a$ becomes
constant (Euclidean static Einstein Universe). For $C\neq 1,0$ the
solutions describe creation of the baby universes (which was invoked
in the Coleman's idea of the ``Big fix''). Remarkably, under the
above conditions, the total action (gravitational plus YM) is
precisely zero~\cite{Donets:1992ck}:\be S_{YM}+S_{gr}=0. \ee Thus,
the pinching off of baby universes occurs with the unit probability.

\section{Non-Abelian Born-Infeld (NBI)} Open string theory suggests
the following generalization of the Maxwell Lagrangian (applicable
to any dimensions):   \be
L=\frac{\beta^2}{4\pi}\left(\sqrt{-\det(g_{\mu\nu}+
F_{\mu\nu}/\beta)}-\sqrt{-g}\right),\ee $\beta$ being the critical
BI field strength ($\beta=1/2\pi \alpha'$ in string theory). In four
dimensions this is equivalent to\be\label{sqrt}
  L=\frac{\beta^2}{4\pi}(\cR-1),\quad
 \cR=\sqrt{1+\frac{F_{\mu\nu}F^{\mu\nu}}{\beta^2}- \frac{(\tilde
F_{\mu\nu}F^{\mu\nu})^2}{16 \beta^4}}.\ee In the non-Abelian case
the strength tensor $F^{\mu\nu}$ is matrix valued, and the
prescription (Tseytlin~\cite{Tseytlin:1997csa}) is more complicated:
the symmetrized trace, which is calculated expanding the Lagrangian
in powers of $F_a^{\mu\nu}T^a$ ($T^a$ being the gauge group
generators), then symmetrizing all products of $T^a$  involved and
only afterwards taking the trace. Symbolically this is given by the
expression \be L_{Str}=\frac{\beta^2}{4\pi} {\mathrm{Str}}
\left(\sqrt{-\det(g_{\mu\nu}+F_{\mu\nu}/\beta)}-
\sqrt{-g}\right),\ee but actually this is a useful form if one is
able to perform a subsequent resummation. Fortunately, this is
possible in the closed from for the homogeneous and isotropic SU(2)
YM field and the metric $ds^2=N^2 dt^2-a^2dl_3^2$ leading
to~\cite{Gal'tsov:2003xm}:\be L_{Str}=- N a^3
\frac{1-2K^2+2V^2-3V^2K^2}{\sqrt{1- K^2+ V^2- K^2V^2}},\quad
 K^2=\frac{\dot{w}^2}{\beta^2 a^2 N^2},\quad
V^2=\frac{(w^2-k)^2}{\beta^2 a^4}.\ee A simpler (the ordinary trace)
prescription for the NBI Lagrangian consists in summation over color
indices in the field invariants $F^a_{\mu\nu}F^{a\mu\nu},\;\; \tilde
F^a_{\mu\nu}F^{a\mu\nu}$ in the square root form of the Lagrangian
(\ref{sqrt}). This gives \be L=- N a^3 \sqrt{1-3 K^2+3 V^2- 9
K^2V^2}.\ee
\subsection{NBI cosmology}  Homogeneous and
isotropic NBI cosmology with an ordinary trace Lagrangian turns out
to be completely solvable by separation of
variables~\cite{Dyadichev:2001su}.  It leads to an interesting
equation of state:\be
p=\frac{\epsilon(\epsilon_c-\epsilon)}{3(\epsilon_c+\epsilon)}, \ee
where $ \epsilon_c=\beta/4\pi$  is the critical energy density,
corresponding to vanishing pressure. For larger energies the
pressure becomes negative, its limiting value being \be
p=-\epsilon/3.\ee This is the equation of state of an ensemble of
non-interacting isotropically distributed straight Nambu-Goto
strings (which indicates on the stringy origin of the NBI
Lagrangian). In the low-energy limit the YM equation of state
$p=\epsilon/3$  is recovered. Thus, the NBI FRW cosmology smoothly
interpolates between the string gas cosmology and the hot Universe.
The energy density is \be\epsilon=\epsilon_c
\;\left(\sqrt{\frac{a^4+3(w^2-k)^2}{a^4-3\dot w^4 }}-1\right).
 \ee

From the YM (NBI) equation one obtains  the following  evolution
equation
 for  the energy density:
\begin{equation}\label{drho}
     \dot {\varepsilon } =-2\frac{ \dot{a}}{a}\,\frac{\varepsilon
     \left (\varepsilon+2 \ec\right )}{\varepsilon+ \ec },
\end{equation}
which can be   integrated to give
\begin{equation}\label{aepsC}
     a^4(\varepsilon+2\ec)\varepsilon={\rm const}.
\end{equation}
 From this relation one can see that the behavior of the NBI
field interpolates between two patterns: 1)~for large energy
densities ($\varepsilon\gg \ec $) the energy density scales as
$\ve\sim a^{-2}$; 2)~for small densities $\varepsilon\ll \ec$  one
has a radiation law $\varepsilon \sim \ a^{-4}$.

Remarkably, the equation for the scale factor $a$ can be decoupled
($g=\beta G$): \be\ddot a = -\frac{2 g a (\dot a ^2+k)}{2 g a^2 + 3
(\dot a^2 + k)}\] and admits the first integral \[3 \left(\dot a^2 +
k \right)^2 + 4 g a^2 \left(\dot a^2 + k \right) = C,\ee which
allows to draw phase portraits for different $k$ \bei\item  {\bf
Closed.}
 The only singular point is $a=0$, $b=0,\;(b=\dot a)$ which is a
center with the eigenvalues $\pm i\sqrt{6g}$.  Solutions evolve from
left to right in the upper half-plane as time changes from~$-\infty$
to~$\infty$, and from right to left in the lower half-plane. All
solutions are of an oscillating type: they start at the singularity
($a=0$) and after a stage of expansion shrink to another
singularity. The global qualitative behavior of  solutions does not
differ substantially from that in the conformally invariant YM field
model, except near the singularity: \be
     a(t) = b_0 t - \frac{b_0 g}{9} t^3
            + O(t^5),
\ee where $b_0$ is a free parameter. Absence of the quadratic term
means that the Universe starts with zero acceleration in accord with
the equation of state  $p\approx-\ve/3$ at high densities.

\item {\bf Spatially flat.}
  There is a singular line
$b=0$ each point of which represents a solution for an empty space
(Minkowski spacetime). This set is degenerate, and there are no
solutions that reach this curve for finite values of~$a$. All
solutions in the upper half-plane after initial singularity expand
infinitely. A remarkable fact is that for this case one can write an
exact solution for $a$ in an implicit form:
\be
     4 \sqrt{g} \, (t - t_0)
     = \sqrt{3} \left(\Omega - \arctan \Omega^{-1} + \pi/2 \right),
\ee where $ \Omega=\sqrt{2}\,a/\sqrt{\sqrt{a^4+C}-a^2} $. The metric
singularity is reached at $t=t_0$.

\item {\bf Open.}
  Physical domain   is $b<-1,\,b>1$. There is a center at $a=0$, $b=0$
with the eigenvalues $\pm i\sqrt{6g}$, but it lies outside the
boundary of the physical region.  Other  singular points are
$(a=0,\;\; b=\pm 1)$. These points are degenerate and cannot be
reached from any point lying in the physically allowed domain of the
phase plane. The only solutions which start from them are the
separatrices $b=\pm 1$ that represent (part of) the flat Minkowski
spacetime in special coordinates. One can easily see that all
solutions in the upper part of the physical domain $\dot a>1$ start
from the singularity and then move to
 $a \to \infty$, $ \dot a \to 1$.
\eei
\subsection{ NBI universe with   $ \Lambda $}
Cosmological constant gives rise to new types of solutions including
de Sitter-like, which are non-singular. However, in the case when
there is a singularity, its structure is determined by the leading
term in equation of state, namely, by the BI pressure, and is
unaffected by $\Lambda$. The generic solution  near the singularity
starts as \be
     a(t)= b_0 t - \frac{2g-\Lambda}{18} b_0 \,t^3 +O(t^5), \ee
with zero acceleration. The equation for the scale factor can be
separated again and admits the first integral \be
     3 \left(b^2-\frac\Lambda3 a^2+k \right)^2
     + 4 g a^2 \left(b^2-\frac\Lambda 3 a^2+k \right)= C,
\ee which gives phase portraits for different k. The allowed domain
now is \vspace{-.2cm}\[\dot a^2 - \frac \Lambda 3 a^2 +k \ge
0\vspace{-.2cm}\] For $k=1$ and a negative cosmological constant the
only singular point is $(a=0,b=0)$, which is a center with
eigenvalues $\xi = \pm 3 i \sqrt{3(2g-\Lambda)}$. All phase space
trajectories are deformed circles, and all solutions start from the
singularity and reach the singularity in the future. For $\Lambda>0$
the allowed domain is bounded by a hyperbola $b^2-\frac \Lambda  3
a^2>-1$ which represents the de Sitter space. Behavior of the
solutions depends on whether $\Lambda$ is smaller or greater than
the quantity $2g$. \bei \item{\bf Closed with $\Lambda$.}
 Physical domains are bounded by hyperbolas. The point
$a=0, b=0$ remains a center. Another pair of singular (saddle)
points $ a =\pm\sqrt{\frac{3(2 g-\Lambda)}{\Lambda (4 g- \Lambda
)}},\; b=0$ lie within  the allowed domain and correspond to the
static Einstein universe.  Entering them separatrices correspond to
either  a solution developing from the singularity into the static
universe, or one rolling down from an infinite radius  to the static
universe. They divide the phase space into the domains containing
different types of generic solutions. Near the origin, all solutions
evolve from an initial to a final singularity. In the region near
the physical boundary, the solutions are non-singular and evolve for
an infinite time first shrinking to some finite value of the scale
factor and then ever expanding. Generic solutions of the third type
possess  a singularity but are non-periodic: the universe expands
forever.

\item
{\bf $\Lambda>2g$ }  With  increasing $\Lambda$, the saddle point
approaches the origin and finally, when $\Lambda>2 g$, swallows it.
The character of the singular point $(a=0,b=0)$ changes---it becomes
a saddle point. The separatrices entering it are solutions of an
inflationary type. They divide all generic solutions into two
classes: nonsingular de Sitter-like   and ever-expanding solutions
with an initial singularity.

\item {\bf Open and flat with $\Lambda$. }  For $k=0$ the only
singular point is the origin $(a=0,b=0)$ which now is degenerate.
For $\Lambda<0$ the allowed domain is the whole plane. All solutions
are of an oscillatory type evolving from the initial to the final
singularity for a finite time.

For a positive  cosmological constant the allowed domain consists of
the upper and lower parts of a cone $b^2 - \frac \Lambda 3 a^2>0$
whose boundary corresponds to the (part of) de Sitter space
described in the inflating coordinates. All solutions that lie
within the allowed domain have an initial singularity and are ever
expanding in the future.

For $k=-1$ the allowed domain lies outside of  either  an ellipse
($\Lambda<0$), or  a hyperbola ($\Lambda>0$). In this case two types
of singular points---the origin and $ a =\pm\sqrt{\frac{3(\Lambda-2
g)}{\Lambda (4g-\Lambda)}},\; b=0 $ (when they are real) lie outside
the allowed domain. Another pair of singular points is $(a=0,b=\pm
1) $. These are degenerate (as in case  with zero cosmological
constant). Since they lie on the boundary of the allowed domain, the
only physical solutions approaching them correspond  to zero energy
of the BI field ((anti)-deSitter).

All  generic solutions within the allowed domain possess a
singularity and are either oscillating (for $\Lambda<0$) or ever
expanding  (for $\Lambda>0$).\eei
 \subsection{Dynamics of YM field}
 The NBI equation for $w$ can be integrated once, giving
 a first order equation \be\frac{\dot{w}^2}{a_0^4-(k-w^2)^2} =\frac{ a^4}{a^4+ 3
 a_0^4},\ee whose solution is oscillatory and  is given by Jacobi
 elliptic functions.
  The main difference with the ordinary EYM cosmology
relates to small values of~$a$. One can  see that near the
singularity ($a\to 0$) the YM oscillations in the NBI case slow
down, while in the ordinary YM cosmology the frequency remains
constant in the conformal gauge. Near the singularity \be
w=w_0+\frac{b_0\alpha}{6(k+b_0^2)} t^2+ O(t^4), \ee where
$\alpha=\pm\sqrt{3(k+b_0^2)^2 - 4 g^2(k-w_0^2)^2}$, $w_0$ is a free
parameter,  $b_0$ is a metric  expansion parameter.

\subsection{NBI on the brane}
 Replacement of the standard YM Lagrangian by the Born-Infeld one breaks
conformal symmetry, providing deviation from the hot equation of
state and creating negative pressure. Surprisingly enough, putting
the same NBI theory into the RS2 framework gives rise to an exact
restoration of the conformal symmetry by the brane non-linear
corrections~\cite{Gal'tsov:2003xm}. Choosing the ordinary trace
action  \be S=\lambda
 {\mathrm{Tr}}\int \sqrt{-\det(g_{\mu\nu}+F_{\mu\nu}/\beta)}\,d^4 x
 - \kappa^2 \int\, (R_5+2\Lambda_5)\sqrt{-g_{5}}\,d^5x,
  \ee
 where the brane tension $\lambda$ plays a role of the BI
 critical energy density, one obtains the constraint equation  \be \left(\frac{\dot{a}}{a}\right)^2=\frac{\kappa^2}{6} \Lambda
  + \frac{\kappa^4}{36}
  (\lambda  +\ve)^2+\frac{\mathcal{E}}{a^4}-\frac{k}{a^2},\ee
  where $\mathcal{E}$ is the integration constant corresponding to the
bulk Weyl tensor projection (``dark radiation'') and, as usual, \be
\Lambda_4 = \frac12 \kappa^2(\Lambda+\frac16 \kappa^2 \lambda ^2),
  \quad  G_{(4)}=\frac{\kappa^4 \lambda  }{48 \pi }.\ee The energy
  density in this model scales as \be \ve = \lambda
  \left(\sqrt{(1+ {C}/{a^4}}-1\right),
\ee where $C$ is the integration constant. Surprisingly, the
constraint equation  comes back to that of the YM conformally
symmetric cosmology \be
  \left(\frac{\dot{a}}{a}\right)^2=\frac{8\pi G_{(4)}}{3}\Lambda_4
  +\frac{\mathcal{C}~}{a^4}-\frac{k}{a^2},\ee
where  the constant $\mathcal {C}= \mathcal{E} + \kappa^4\lambda^2
C/36$ includes contributions from both the ``dark radiation'' and
the YM energy density.

\subsection{ Smoothing YM chaos with NBI}  In flat space
the homogeneous YM dynamics is typically chaotic
\cite{Matinyan:1981dj}. When gravity is switched on one deals with
cosmological solutions. The occurrence of chaos is then related to
additional symmetries: in the isotropic cosmology case, as we have
seen, the evolution is perfectly regular, but once an isotropy is
broken, the chaotic behavior is typically manifest
(see~\cite{Darian:1996jf,Barrow:1997sb} for Bianchi~I
and~\cite{Jin:2004vh} for other
 anisotropic Bianchi types).  When the quadratic YM action is changed
 to the NBI form, it turns out  that YM evolution becomes more regular.
This may be interpreted as  smoothing stringy effect on classical
chaos~\cite{Gal'tsov:2003gx,Dyadichev:2004ix,Dyadichev:2006xc}. For
the Bianchi I metric \be ds^2=N^2dt^2-b^2(dx^2+dy^2)-c^2dz^2,\ee
with $N$, $b$ and $c$ depending on time,   the corresponding YM
field is parameterized two functions $u, v$ of time: $ A=T_1 u dx
+T_2 u dy + T_3 v
  dz. \;\;$
The   Poincare sections signal the chaos-order transition at
$\beta_{\rm cr}=0.317$ (regular from the strong NBI side
$\beta<\beta_{\rm cr}$).

\section{Conformal symmetry breaking and DE}  Conformal symmetry
breaking in NBI theory demonstrates the occurrence of the negative
pressure, but its extremal value $p=-\ve/3$ is still insufficient
for DE. Meanwhile, a stronger violation of conformal symmetry may
provide an equation of state with $\ve\sim -1$. Such violation can
be of different nature: \bei \item Quantum corrections,
\item Non-minimal coupling to gravity, \item Dilaton and other
coupled scalar fields including Higgs, \item String theory
corrections.\eei Here we just explore some model Lagrangians to see
the necessary conditions for DE. Assuming the Lagrangian to be an
arbitrary function $L(\cF,\cG)$ of invariants \be
\cF=-F^a_{\mu\nu}F^{a\mu\nu}/2\;\; {\rm and}\;\; \cG=-\tilde
F^a_{\mu\nu}F^{a\mu\nu}/4,\ee one finds for the pressure and the
energy density (conformal time):  \bea p&=&L +\left( 2\frac{\partial
L}{\partial \cF} [2(k-w^2)^2-\dot w^2]- 3\frac{\partial L}{\partial
\cG}\dot w(k-w^2)\right)a^{-4}, \non\\ \vspace{.5cm} \ve&=&-L
+\left( 6\frac{\partial L}{\partial \cF} \dot w^2+3\frac{\partial
L}{\partial \cF}\dot w(k-w^2)\right)a^{-4}.\eea
 For a simple estimate consider the power-low
dependence: \be L\sim \cF (\cF/\mu^2)^{\nu-1},\ee where $\mu$ has
the dimension of mass. Then in the electric (kinetic) dominance
regime one obtains \be W=\frac{p}{\ve}=\frac{3-2\nu}{3(2\nu-1)}.\ee
For certain $\nu$ this quantity may be arbitrarily close to $W=-1$
or even less. An electric phantom regime is thus possible.

In the magnetic (potential) dominance  regime one obtains \be
W=4\nu/3-1.\ee The value $W=-1$ can not be reached, but an
admissible DE regime is also possible. These regimes are transient
since during the evolution the electric part transforms to the
magnetic and vice versa.

Another plausible form of the lagrangian (suggested by quantum
corrections) is logarithmic~\cite{Zhao:2005bu}: \be L\sim \cF
\ln(\cF/\mu^2).\ee Then the energy density is \be \ve=3\left(
T[\ln(\cF/\mu^2)+2]+V \ln(\cF/\mu^2)\right)a^{-4}, \ee and the
equation of state is \be
W=\frac{p}{\ve}=\frac{(T+V)\ln(\cF/\mu^2)+2(2V-T)}
{3(T+V)\ln(\cF/\mu^2)+6T},\ee where $T=\dot w^2,\; V=(k-w^2)^2.$ It
is easy to see that $W\sim-1$ for $\ln(\cF/\mu^2)\sim -1$. In this
case the DE regime is transient. The phantom regime is also
possible.

Thus, the DE conditions for the homogeneous and isotropic YM field
can arise indeed as a result of sufficiently strong breaking of the
conformal symmetry.

\section{Outlook} Attractive features of the
hypothesis of the non-Abelian condensates as dark energy are clear
physical meaning of the Yang-Mills field, the existence of the
natural isotropic and homogeneous configuration removing the problem
of isotropy of Abelian vector field cosmological models, as well as
some non-trivial effects related to the YM topology. The major yet
unsolved problem is the mechanism  of the conformal symmetry
breaking which is necessary to ensure an accelerating expansion.
Various ideas for this were suggested, but further work is needed to
support or reject the hypothesis. Also, physical mechanisms of
generation of non-Abelian condensates in cosmology require more
detailed investigation.

\section*{Acknowledgments}
The author would like to thank the Organizing Committee  for
invitation and support and especially Jean Tran Thanh Van for a warm
and stimulating atmosphere of the conference. The work was
supported by RFBR under projects 08-02-01398, 08-02-08066.


\section*{References}
\begin{thebibliography}{99}

\bibitem{ArmendarizPicon:2004pm}
  C.~Armendariz-Picon,
  ``Could dark energy be vector-like?,''
  JCAP {\bf 0407}, 007 (2004)
  [arXiv:astro-ph/0405267].

\bibitem{Kiselev:2004py}
  V.~V.~Kiselev,
  ``Vector field as a quintessence partner,''
  Class.\ Quant.\ Grav.\  {\bf 21}, 3323 (2004)
  [arXiv:gr-qc/0402095].

\bibitem{Zhao:2005bu}
  W.~Zhao and Y.~Zhang,
  ``The state equation of the Yang-Mills field dark energy models,''
  Class.\ Quant.\ Grav.\  {\bf 23}, 3405 (2006)
  [arXiv:astro-ph/0510356].

\bibitem{Wei:2006tn}
  H.~Wei and R.~G.~Cai,
  ``Interacting vector-like dark energy, the first and second cosmological
  Phys.\ Rev.\  D {\bf 73}, 083002 (2006)
  [arXiv:astro-ph/0603052].

\bibitem{Fuzfa:2006pn}
  A.~Fuzfa and J.~M.~Alimi,
  ``Dark energy as a Born-Infeld gauge interaction violating the  equivalence
  Phys.\ Rev.\ Lett.\  {\bf 97}, 061301 (2006)
  [arXiv:astro-ph/0604517].

\bibitem{Zhao:2007vn}
  W.~Zhao, D. Xu,
  ``Evolution of magnetic component in Yang-Mills condensate dark energy
  Int.\ J.\ Mod.\ Phys.\  D {\bf 16}, 1735 (2007)
  [arXiv:gr-qc/0701136].

\bibitem{Jimenez:2008au}
  J.~B.~Jimenez and A.~L.~Maroto,
  ``A cosmic vector for dark energy,''
  arXiv:0801.1486 [astro-ph].


\bibitem{Bamba:2008xa}
  K.~Bamba, S.~Nojiri and S.~D.~Odintsov,
  ``Inflationary cosmology and the late-time accelerated expansion of the
  arXiv:0803.3384 [hep-th].

\bibitem{Copeland:2006wr}
  E.~J.~Copeland, M.~Sami and S.~Tsujikawa,
  ``Dynamics of dark energy,''
  Int.\ J.\ Mod.\ Phys.\  D {\bf 15}, 1753 (2006)
  [arXiv:hep-th/0603057].

\bibitem{Linde:1979pr}
  A.~D.~Linde,
  ``Classical Yang-Mills Solutions, Condensation Of W Mesons And Symmetry Of
  Phys.\ Lett.\  B {\bf 86}, 39 (1979).

\bibitem{Ford:1989me}
  L.~H.~Ford,
  ``INFLATION DRIVEN BY A VECTOR FIELD,''
  Phys.\ Rev.\  D {\bf 40}, 967 (1989).

\bibitem{Zimdahl:2000zm}
  W.~Zimdahl, D.~J.~Schwarz, A.~B.~Balakin and D.~Pavon,
  ``Cosmic anti-friction and accelerated expansion,''
  Phys.\ Rev.\  D {\bf 64}, 063501 (2001)
  [arXiv:astro-ph/0009353].

\bibitem{Galtsov:1991un}
  D.~V.~Galtsov and M.~S.~Volkov,
  ``Yang-Mills cosmology: Cold matter for a hot universe,''
  Phys.\ Lett.\  B {\bf 256}, 17 (1991).

\bibitem{Cervero:1978db}
  J.~Cervero and L.~Jacobs,
  ``Classical Yang-Mills Fields In A Robertson-Walker Universe,''
  Phys.\ Lett.\  B {\bf 78}, 427 (1978).

\bibitem{Henneaux:1982vs}
  M.~Henneaux,
  ``Remarks On Space-Time Symmetries And Nonabelian Gauge Fields,''
  J.\ Math.\ Phys.\  {\bf 23}, 830 (1982).

\bibitem{Hosotani:1984wj}
  Y.~Hosotani,
  ``Exact Solution To The Einstein Yang-Mills Equation,''
  Phys.\ Lett.\  B {\bf 147}, 44 (1984).

\bibitem{Bertolami:1990je}
  O.~Bertolami, J.~M.~Mourao, R.~F.~Picken and I.~P.~Volobuev,
  ``Dynamics of euclidenized Einstein Yang-Mills systems with arbitrary gauge
  Int.\ J.\ Mod.\ Phys.\  A {\bf 6}, 4149 (1991).

\bibitem{Moniz:1990hf}
  P.~V.~Moniz and J.~M.~Mourao,
  ``Homogeneous and isotropic closed cosmologies with a gauge sector,''
  Class.\ Quant.\ Grav.\  {\bf 8}, 1815 (1991).



\bibitem{Moniz:1991kx}
  P.~V.~Moniz, J.~M.~Mourao and P.~M.~Sa,
  ``The Dynamics Of A Flat Friedmann-Robertson-Walker Inflationary Model In The
  Class.\ Quant.\ Grav.\  {\bf 10}, 517 (1993).

\bibitem{Kuenzle:1991wa}
  H.~P.~Kuenzle,
  ``SU(n) Einstein Yang-Mills fields with spherical symmetry,''
  Class.\ Quant.\ Grav.\  {\bf 8}, 2283 (1991).


\bibitem{Darian:1996mb}
  B.~K.~Darian and H.~P.~Kunzle,
  ``Cosmological Einstein-Yang-Mills equations,''
  J.\ Math.\ Phys.\  {\bf 38}, 4696 (1997)
  [arXiv:gr-qc/9610026].

\bibitem{Bertolami:1990iw}
  O.~Bertolami and J.~M.~Mourao,
  ``Euclideanized Einstein Yang-Mills Equations, Wormholes And The Ground State
{\it  In Lisbon 1990, Proceedings, The physical universe* 21-38
(QB981:A9:1990).}

\bibitem{Donets:1992ck}
  E.~E.~Donets and D.~V.~Galtsov,
  ``Continuous family of Einstein Yang-Mills wormholes,''
  Phys.\ Lett.\  B {\bf 294}, 44 (1992)
  [arXiv:gr-qc/9209008].

\bibitem{Donets:1992vx}
  E.~E.~Donets and D.~V.~Galtsov,
  ``Wormhole solutions in coupled Einstein Yang-Mills axion system,''
{\it  In *Evora 1992, Proceedings, Classical and quantum gravity*
289-292.}

\bibitem{Cavaglia:1993en}
  M.~Cavaglia and V.~de Alfaro,
  ``On a quantum universe filled with Yang-Mills radiation,''
  Mod.\ Phys.\ Lett.\  A {\bf 9}, 569 (1994)
  [arXiv:gr-qc/9310001].

\bibitem{Bertolami:1994jn}
  O.~Bertolami and P.~V.~Moniz,
  ``Decoherence of Friedmann-Robertson-Walker geometries in the presence of
  Nucl.\ Phys.\  B {\bf 439}, 259 (1995)
  [arXiv:gr-qc/9410027].

\bibitem{Cavaglia:1994zv}
  M.~Cavaglia, V.~De Alfaro and A.~T.~Filippov,
  ``Quantization of the Robertson-Walker Universe,''
{\it  in Proc. Quantum Systems: New Trends And Methods (QS 94) 23-29
May 1994, Minsk, Belarus, pp. 31-46}

\bibitem{Bertolami:1991ff}
  O.~Bertolami, Yu.~A.~Kubyshin and J.~M.~Mourao,
  ``Stability of compactification in Einstein Yang-Mills theories after
  Phys.\ Rev.\  D {\bf 45}, 3405 (1992).

\bibitem{Bento:1994dw}
  M.~C.~Bento and O.~Bertolami,
  ``General Cosmological Features Of The Einstein Yang-Mills Dilaton System In
  Phys.\ Lett.\  B {\bf 336}, 6 (1994)
  [arXiv:gr-qc/9405038].

\bibitem{Moniz:1996ja}
  P.~V.~Moniz,
  ``Quantization of a Friedmann-Robertson-Walker Model with Gauge Fields in N=1
  arXiv:gr-qc/9604045;
  ``FRW model with vector fields in N=1 supergravity,''
  Helv.\ Phys.\ Acta {\bf 69}, 293 (1996).

\bibitem{Fuzfa:2003gc}
  A.~Fuzfa,
  ``Gravitational instability of Yang-Mills cosmologies,''
  Class.\ Quant.\ Grav.\  {\bf 20}, 4753 (2003)
  [arXiv:gr-qc/0310032].

\bibitem{Darian:1996jf}
  B.~K.~Darian and H.~P.~Kunzle,
  ``Axially symmetric Bianchi I Yang-Mills cosmology as a dynamical system,''
  Class.\ Quant.\ Grav.\  {\bf 13}, 2651 (1996)
  [arXiv:gr-qc/9608024].

\bibitem{Barrow:1997sb}
  J.~D.~Barrow and J.~J.~Levin,
  ``Chaos in the Einstein-Yang-Mills equations,''
  Phys.\ Rev.\ Lett.\  {\bf 80}, 656 (1998)
  [arXiv:gr-qc/9706065].

\bibitem{Jin:2004vh}
  Y.~Jin and K.~i.~Maeda,
  ``Chaos of Yang-Mills field in class A Bianchi spacetimes,''
  Phys.\ Rev.\  D {\bf 71}, 064007 (2005)
  [arXiv:gr-qc/0412060].

\bibitem{Matinyan:1981dj}
  S.~G.~Matinyan, G.~K.~Savvidy and N.~G.~Ter-Arutunian Savvidy,
  ``Classical Yang-Mills Mechanics. Nonlinear Color Oscillations,''
  Sov.\ Phys.\ JETP {\bf 53}, 421 (1981)
  [Zh.\ Eksp.\ Teor.\ Fiz.\  {\bf 80}, 830 (1981)].

\bibitem{Donets:1996ja}
  E.~E.~Donets, D.~V.~Galtsov and M.~Y.~Zotov,
  ``Internal structure of Einstein Yang-Mills black holes,''
  Phys.\ Rev.\  D {\bf 56}, 3459 (1997)
  [arXiv:gr-qc/9612067].

\bibitem{Donets:1998xd}
  E.~E.~Donets, M.~N.~Tentyukov and M.~M.~Tsulaia,
  ``Evolution of nonlinear perturbations inside Einstein-Yang-Mills black
  Phys.\ Rev.\  D {\bf 59}, 064008 (1999)
  [arXiv:gr-qc/9808025].

\bibitem{Gal'tsov:1997uq}
  D.~V.~Gal'tsov, E.~E.~Donets and M.~Y.~Zotov,
  ``Singularities inside hairy black holes,''
  arXiv:gr-qc/9712003.

\bibitem{Galtsov:1997iy}
  D.~V.~Galtsov, E.~E.~Donets and M.~Y.~Zotov,
  ``Internal structure of non-Abelian black holes,''
  In *Haifa 1997, Internal structure of black holes and spacetime singularities* 142-162.
  arXiv:hep-th/9709181.

\bibitem{Galtsov:1997ub}
  D.~V.~Galtsov and E.~E.~Donets,
``Power-law mass inflation in Einstein-Yang-Mills-Higgs black
holes,''
 Comptes Rend. Acad. Sci. Paris, t.325, Serie IIB (1997) 649-657,
  arXiv:gr-qc/9706067.

\bibitem{Galtsov:1991du}
  D.~V.~Galtsov and M.~S.~Volkov,
  ``Sphalerons in Einstein Yang-Mills theory,''
  Phys.\ Lett.\  B {\bf 273}, 255 (1991).

\bibitem{Donets:1992zb}
  E.~E.~Donets and D.~V.~Galtsov,
  ``Stringy sphalerons and nonAbelian black holes,''
  Phys.\ Lett.\  B {\bf 302}, 411 (1993)
  [arXiv:hep-th/9212153].

\bibitem{Ding:1995jf}
  S.~X.~Ding,
 ``Evolution and fate of perturbed Bartnik-McKinnon spacetime,''
Phys.\ Lett.\  B {\bf 381}, 49 (1996).

\bibitem{Volkov:1998cc}
  M.~S.~Volkov and D.~V.~Gal'tsov,
  ``Gravitating non-Abelian solitons and black holes with Yang-Mills  fields,''
  Phys.\ Rept.\  {\bf 319}, 1 (1999)
  [arXiv:hep-th/9810070].

\bibitem{Gal'tsov:2001tx}
  D.~V.~Gal'tsov,
  ``Gravitating lumps,''
  in {\it General Relativity and
Gravitation, Proc. of the 16th International conference}, Durban,
South Africa, 15-21 July 2001, ed. N.T. Bishop and S.D. Maharaj,
World Scientific, 2002, pp. 142-161;
  arXiv:hep-th/0112038.

\bibitem{Tipler:2007vx}
  F.~J.~Tipler,
  ``Feynman-Weinberg Quantum Gravity and the Extended Standard Model as a
  Rept.\ Prog.\ Phys.\  {\bf 68}, 897 (2005)
  [arXiv:0704.3276 [hep-th]].

\bibitem{Tseytlin:1997csa}
  A.~A.~Tseytlin,
  ``On non-abelian generalisation of the Born-Infeld action in string
  Nucl.\ Phys.\  B {\bf 501}, 41 (1997)
  [arXiv:hep-th/9701125].

\bibitem{Gal'tsov:1999vn}
  D.~Gal'tsov and R.~Kerner,
  ``Classical glueballs in non-Abelian Born-Infeld theory,''
  Phys.\ Rev.\ Lett.\  {\bf 84}, 5955 (2000)
  [arXiv:hep-th/9910171].

\bibitem{Dyadichev:2001su}
  V.~V.~Dyadichev, D.~V.~Gal'tsov, A.~G.~Zorin and M.~Y.~Zotov,
  ``Non-Abelian Born-Infeld cosmology,''
  Phys.\ Rev.\  D {\bf 65}, 084007 (2002)
  [arXiv:hep-th/0111099].

\bibitem{Moniz:2002rd}
  P.~V.~Moniz,
  ``Frw Quantum Cosmology In The Nonabelian Born-Infeld Theory,''
  Class.\ Quant.\ Grav.\  {\bf 19}, L127 (2002).

\bibitem{Gal'tsov:2003xm}
  D.~V.~Gal'tsov and V.~V.~Dyadichev,
  ``Non-Abelian brane cosmology,''
  Astrophys.\ Space Sci.\  {\bf 283}, 667 (2003)
  [arXiv:hep-th/0301044].

\bibitem{Fuzfa:2005qn}
  A.~Fuzfa and J.~M.~Alimi,
  ``Non-Abelian Einstein-Born-Infeld dilaton cosmology,''
  Phys.\ Rev.\  D {\bf 73}, 023520 (2006)
  [arXiv:gr-qc/0511090].

\bibitem{Novello:2006ng}
  M.~Novello, E.~Goulart, J.~M.~Salim and S.~E.~Perez Bergliaffa,
  ``Cosmological effects of nonlinear electrodynamics,''
  Class.\ Quant.\ Grav.\  {\bf 24}, 3021 (2007)
  [arXiv:gr-qc/0610043].

\bibitem{Gal'tsov:2003gx}
  D.~V.~Gal'tsov and V.~V.~Dyadichev,
  ``Stabilization of the Yang-Mills chaos in non-Abelian Born-Infeld theory,''
  JETP Lett.\  {\bf 77}, 154 (2003)
  [Pisma Zh.\ Eksp.\ Teor.\ Fiz.\  {\bf 77}, 184 (2003)]
  [arXiv:hep-th/0301069].

\bibitem{Gal'tsov:2003gx}
  D.~V.~Gal'tsov and V.~V.~Dyadichev,
  ``Stabilization of the Yang-Mills chaos in non-Abelian Born-Infeld theory,''
  JETP Lett.\  {\bf 77}, 154 (2003)
  [Pisma Zh.\ Eksp.\ Teor.\ Fiz.\  {\bf 77}, 184 (2003)]
  [arXiv:hep-th/0301069].

\bibitem{Dyadichev:2004ix}
  V.~V.~Dyadichev, D.~V.~Gal'tsov and P.~Vargas Moniz,
  ``Chaos - order transition in Bianchi I non-Abelian Born-Infeld  cosmology,''
  Phys.\ Rev.\  D {\bf 72}, 084021 (2005)
  [arXiv:hep-th/0412334].

\bibitem{Dyadichev:2006xc}
  V.~V.~Dyadichev, D.~V.~Galtsov and P.~V.~Moniz,
  ``New features about chaos in Bianchi I non-Abelian Born-Infeld cosmology,''
  AIP Conf.\ Proc.\  {\bf 861}, 312 (2006).

\bibitem{Gibbons:1993pq}
  G.~W.~Gibbons and A.~R.~Steif,
  ``Yang-Mills cosmologies and collapsing gravitational sphalerons,''
  Phys.\ Lett.\  B {\bf 320}, 245 (1994)
  [arXiv:hep-th/9311098].

\bibitem{Bartnik:1988am}
  R.~Bartnik and J.~Mckinnon,
  ``Particle - Like Solutions of the Einstein Yang-Mills Equations,''
  Phys.\ Rev.\ Lett.\  {\bf 61}, 141 (1988).

\bibitem{Volkov:1993gp}
  M.~S.~Volkov,
  ``Einstein Yang-Mills sphalerons and fermion number nonconservation,''
  Phys.\ Lett.\  B {\bf 328}, 89 (1994)
  [arXiv:hep-th/9312005].

\bibitem{Volkov:1996hm}
  M.~S.~Volkov,
  ``Computation of the winding number diffusion rate due to the cosmological
  Phys.\ Rev.\  D {\bf 54}, 5014 (1996)
  [arXiv:hep-th/9604054].

\bibitem{Ding:1994nw}
  S.~X.~Ding,
  ``Cosmological sphaleron from real tunneling and its fate,''
  Phys.\ Rev.\  D {\bf 50}, 3755 (1994)
  [arXiv:gr-qc/9407036].

\bibitem{Hosoya:1989zn}
  A.~Hosoya and W.~Ogura,
  ``Wormhole Instanton Solution In The Einstein Yang-Mills System,''
  Phys.\ Lett.\  B {\bf 225}, 117 (1989).

\bibitem{Das:1989ne}
  A.~K.~Das and J.~Maharana,
  ``Wormhole solution in coupled Yang-Mills axion system,''
  Phys.\ Rev.\  D {\bf 41}, 699 (1990).

\bibitem{Rey:1989th}
  S.~J.~Rey,
  ``SPACE-TIME WORMHOLES WITH YANG-MILLS FIELDS,''
  Nucl.\ Phys.\  B {\bf 336}, 146 (1990).

\bibitem{Gupta:1989bs}
  A.~K.~Gupta, J.~Hughes, J.~Preskill and M.~B.~Wise,
  ``Magnetic Wormholes And Topological Symmetry,''
  Nucl.\ Phys.\  B {\bf 333}, 195 (1990).

\bibitem{Verbin:1989sg}
  Y.~Verbin and A.~Davidson,
  ``Quantized Nonabelian Wormholes,''
  Phys.\ Lett.\  B {\bf 229}, 364 (1989).

\end{thebibliography}
\end{document}